\documentclass[11pt,letterpaper]{article}
\usepackage{emnlp2017}
\usepackage{times}
\usepackage{latexsym}
\usepackage{subcaption}
\usepackage{graphicx}
\usepackage{fancybox}
\usepackage{float}
\emnlpfinalcopy

\def\emnlppaperid{***}

\title{An Entity Resolution Approach to Isolate Instances of\\ Human Trafficking Online}

\author{Chirag Nagpal, Kyle Miller, Benedikt Boecking \and Artur Dubrawski\\
  {\small \tt chiragn@cs.cmu.edu, mille856@andrew.cmu.edu, bboecking@andrew.cmu.edu, awd@cs.cmu.edu}\\
  {Carnegie Mellon University}}

\date{}

\begin{document}

\maketitle

\begin{abstract}
Human trafficking is a challenging law enforcement problem, and a large amount of such activity manifests itself on various online forums. Given the large, heterogeneous and noisy structure of this data, building models to predict instances of trafficking is an even more convolved a task. In this paper we propose and entity resolution pipeline using a notion of proxy labels, in order to extract clusters from this data with prior history of human trafficking activity. We apply this pipeline to 5M records from backpage.com and report on the performance of this approach, challenges in terms of scalability, and some significant domain specific characteristics of our resolved entities.
\end{abstract}

\section{Introduction}

{O}{ver} the years human trafficking has grown to be a challenging law enforcement problem. The advent of the internet has brought the problem in the public domain making it an ever greater societal concern. Prior studies \cite{kennedy2012predictive} have leveraged computational techniques to this data to detect spatio-temporal patterns, by utilizing certain features of the ads. Certain studies \cite{dubrawski2015leveraging} have utilized machine learning approaches to identify if ads could be possibly involved in human trafficking activity. Significant work has also been carried out in building large distributed systems, to store and process such data, and carry out entity resolution to establish ontological relationships between various entities. \cite{szekely15-iswc}

In this paper we explore the possibility of leveraging this information to identify sources of these advertisements, isolate such clusters and identify potential sources of human trafficking from this data using prior domain knowledge. 

In case of ordinary Entity Resolution schemes, each record is considered to represent a single entity. A popular approach in such scenarios is a `merge and purge' strategy whereas records are compared and matched, they are merged into a single more informative record, and the individual records are deleted from the dataset. \cite{benjelloun2009swoosh}

While our problem can be considered a case of Entity Resolution, however, escort advertisements are a challenging, noisy and unstructured dataset. In case of escort advertisements, a single advertisement, may represent one or a group of entities. The advertisements hence might contain features belonging to more than one individual or group. 

The advertisements are also associated with multiple features, including Text, Hyperlinks, Images, Timestamps, Locations etc. In order to featurize characteristics from text we use the regex based information extractor based on the GATE framework \cite{cunningham2002gate}.  This allows us to generate certain domain specific features from our dataset, including, the aliases, cost, location, phone numbers, specific URLs, etc of the entities advertised. We use these features, along with other generic text, the images, etc as features for our classifier. The high reuse of similar features makes it difficult to use exact match over a single feature in order to perform entity resolution. 

\begin{figure*}[!htbp]
\centering
\begin{subfigure}{0.65\textwidth}
\centering
    \includegraphics[width=\linewidth]{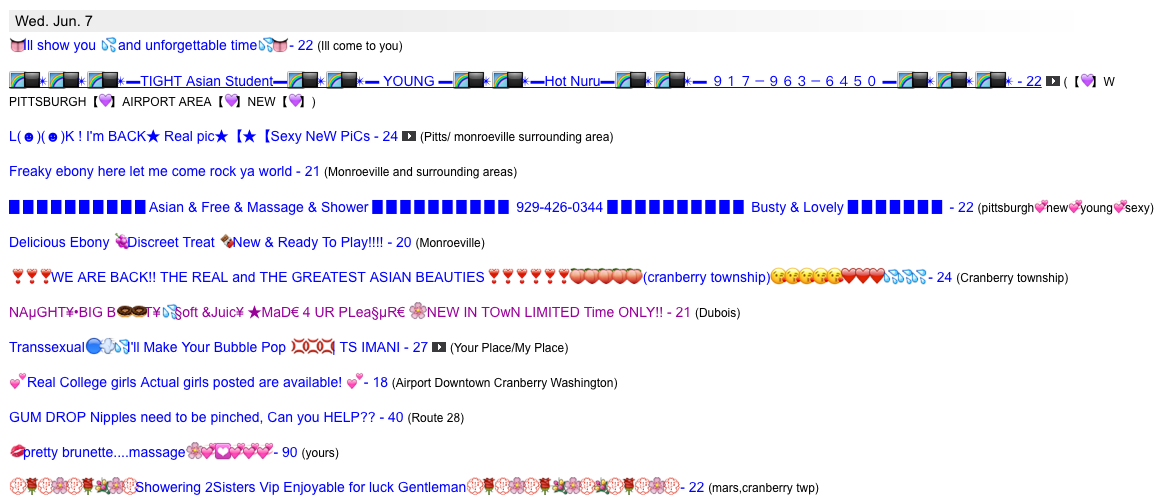}
    \caption{Search Results on \textbf{backpage.com}}
    \label{fig:inclu}
\end{subfigure}%
\begin{subfigure}{0.35\textwidth}
\centering
    \includegraphics[width=0.95\linewidth]{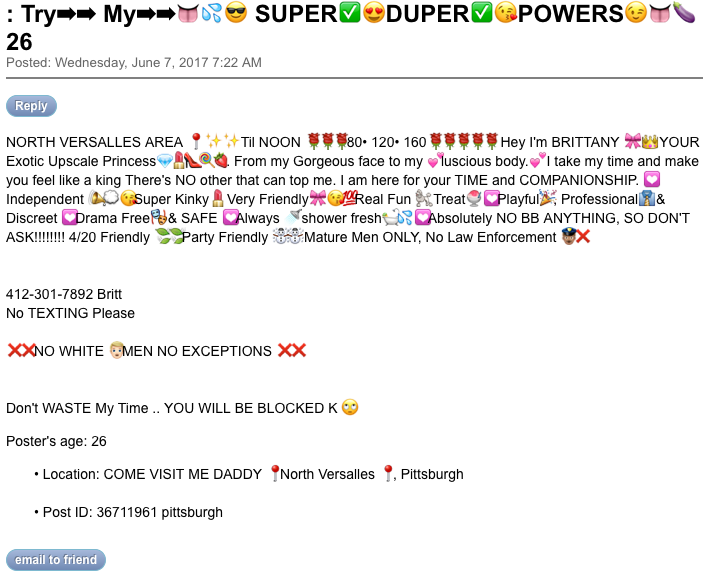}
    \caption{Representative escort advertisement}
    \label{fig:deform}
\end{subfigure}
\caption{Escort advertisements are a classic source of what can be described as Noisy Text. Notice the excessive use of Emojis, Intentional misspelling and relatively benign colloquialisms to obfuscate a more nefarious intent. Domain experts extract meaningful cues from the spatial and temporal indicators, and other linguistic markers to suspect trafficking activity,  which further motivate the leveraging of computational approaches to support such decision making.}
\label{fig:example}
\end{figure*}

We proceed to leverage machine learning approaches to learn a function that can predict if two advertisements are from the same source. The challenge with this is that we have no prior knowledge of the source of advertisements. We thus depend upon a strong feature, in our case Phone Numbers, which can be used as proxy evidence for the source of the advertisements and can help us generate labels for the Training and Test data for a classifier. We can therefore use such strong evidence as to learn another function, which can help us generate labels for our dataset, this semi-supervised approach is described as `surrogate learning' in \cite{veeramachaneni2009surrogate}. Pairwise comparisons result in an extremely high number of comparisons over the entire dataset. In order to reduce this, we use a blocking scheme using certain features. 

\begin{figure*}[!t]
\includegraphics[width=\linewidth]{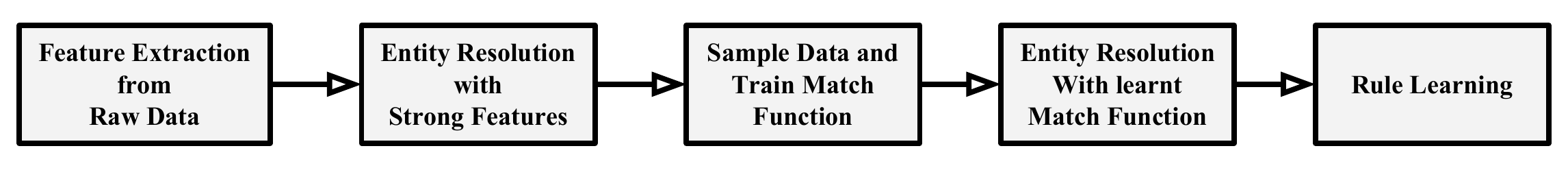}
\caption{The proposed Entity Resolution pipeline }
\label{fig:pipeline}
\end{figure*}

The resulting clusters are isolated for human trafficking using prior expert knowledge and featurized. Rule learning is used to establish differences between these and other components. The entire pipeline is represented by Figure \ref{fig:pipeline}.

\section{Domain and Feature Extraction}
Figure \ref{fig:example} is illustrative of the search results of escort advertisements and a page advertising a particular individual. The text is inundated with special characters, Emojis, as well as misspelled words that are specific markers and highly informative to domain experts. the text consists of information, regarding the escorts area of operation, phone number, any particular client preferences, and the advertised cost. We proceed to build Regular expression based feature extractors to extract this information and store in a fixed schema, using the popular JAPE tool part of the GATE suite of NLP tools. The extractor we build for this domain, \textbf{AnonymousExtractor} is open source and publically available at \url{github.com/mille856/CMU_memex}.
\begin{table}
\caption{Performance of \textbf{TJBatchExtractor}}
\begin{tabular}{l|c|c|c}
\hline
\textbf{Feature}&\textbf{Precision}&\textbf{Recall}&\textbf{$F_{1}$ Score}\\ \hline 
Age&0.980&0.731&0.838 \\
Cost&0.889&0.966&0.926\\
E-mail&1.000&1.000&1.000\\
Ethnicity&0.969&0.876&0.920\\
Eye Color&1.000&0.962&0.981\\
Hair Color&0.981&0.959&0.970\\
Name&0.896&0.801&0.846\\
Phone Number&0.998&0.995&0.997\\
Restriction(s)&0.949&0.812&0.875\\
Skin Color&0.971&0.971&0.971\\
URL&0.854&0.872&0.863\\
Height&0.978&0.962&0.970\\
Measurement&0.919&0.883&0.901\\
Weight&0.976&0.912&0.943\\ \hline
\end{tabular}
\label{tab:tjbe}
\end{table}

Table \ref{tab:tjbe} lists the performance of our extraction tool on 1,000 randomly sampled escort advertisements, for the various features. Most of the features are self explanatory.  (The reader is directed to \cite{dubrawski2015leveraging} for a complete description of the fields extracted.) The noisy nature, along with intentional obfuscations, especially in case of features like Names results in lower performance as compared to the other extracted features.

Apart from the Regular Expression based features, we also extract the hashcodes of the images in the advertisements, the posting date and time, and location.\footnote{These features are present as metadata, and do not require the use of hand engineered Regexs.}

\section{Entity Resolution}

\subsection{Definition}

We approach the problem of extracting connected components from our dataset using pairwise entity resolution. The similarity or connection between two nodes is treated as a learning problem, with training data for the problem generated by using `proxy' labels from existing evidence of connectivity from strong features. 

More formally the problem can be considered to be to sample all connected components $\mathcal{H}_{i}(\mathcal{V}, \mathcal{E})$ from a graph $\mathcal{G}(\mathcal{V},\mathcal{E})$. Here, $\mathcal{V}$, the set of vertices ($\{v_{1}, v_{2}, ..., v_{n} \}$) is the set of advertisements and $\mathcal{E}$, $\{(v_{i}, v_{j}),(v_{j},v_{k}), ..., (v_{k}, v_{l})  \}$ is the set of edges between individual records, the presence of which indicates they represent the same entity.

We need to learn a function $M(v_{i}, v_{j})$ such that $M(v_{i}, v_{j}) = \mathrm{Pr}( (v_{i}, v_{j}) \in \mathcal{E}(\mathcal{H}_{i}), \forall \mathcal{H}_{i} \in \mathcal{H})$

The set of strong features present in a given record can be considered to be the function `$\mathcal{S}$'. Thus, in our problem, $\mathcal{S}_{v}$ represents all the phone numbers associated with $v$.

Thus $\mathcal{S} = \bigcup \mathcal{S}_{v_i}, \forall v_{i} \in \mathcal{V}$. Here, $|\mathcal{S}| << |\mathcal{V}|$

Now, let us further consider the graph $\mathcal{G}^{*}(\mathcal{V}, \mathcal{E})$ defined on the set of vertices $\mathcal{V}$, such that $(v_{i}, v_{j}) \in \mathcal{E}(\mathcal{G}^{*})$ if $|\mathcal{S}_{v_{i}} \cap \mathcal{S}_{v_{j}}|>0$ (more simply, the graph described by strong features.)

Let $\mathcal{H}^{*}$ be the set of all the of connected components $\{ \mathcal{H}^{*}_{1}(\mathcal{V},\mathcal{E}), \mathcal{H}^{*}_{2}(\mathcal{V},\mathcal{E}), ... ,\mathcal{H}^{*}_{n}(\mathcal{V},\mathcal{E}) \}$ defined on the graph $\mathcal{G}^{*}(\mathcal{V},\mathcal{E})$

Now, function $\mathcal{P}$ is such that for any $p_{i} \in \mathcal{S}$%

\noindent $\mathcal{P}(p_{i})= \mathcal{V}(\mathcal{H}^{*}_{k}) \iff p_{i} \in \bigcup \mathcal{S}_{v_i}, \forall v_{i} \in \mathcal{V}(\mathcal{H}^{*}_{k})$



\begin{figure}[!htbp]
\centering
\includegraphics[width=0.75\linewidth, trim={1.9cm 18.1cm 2.3cm 2.6cm},clip]{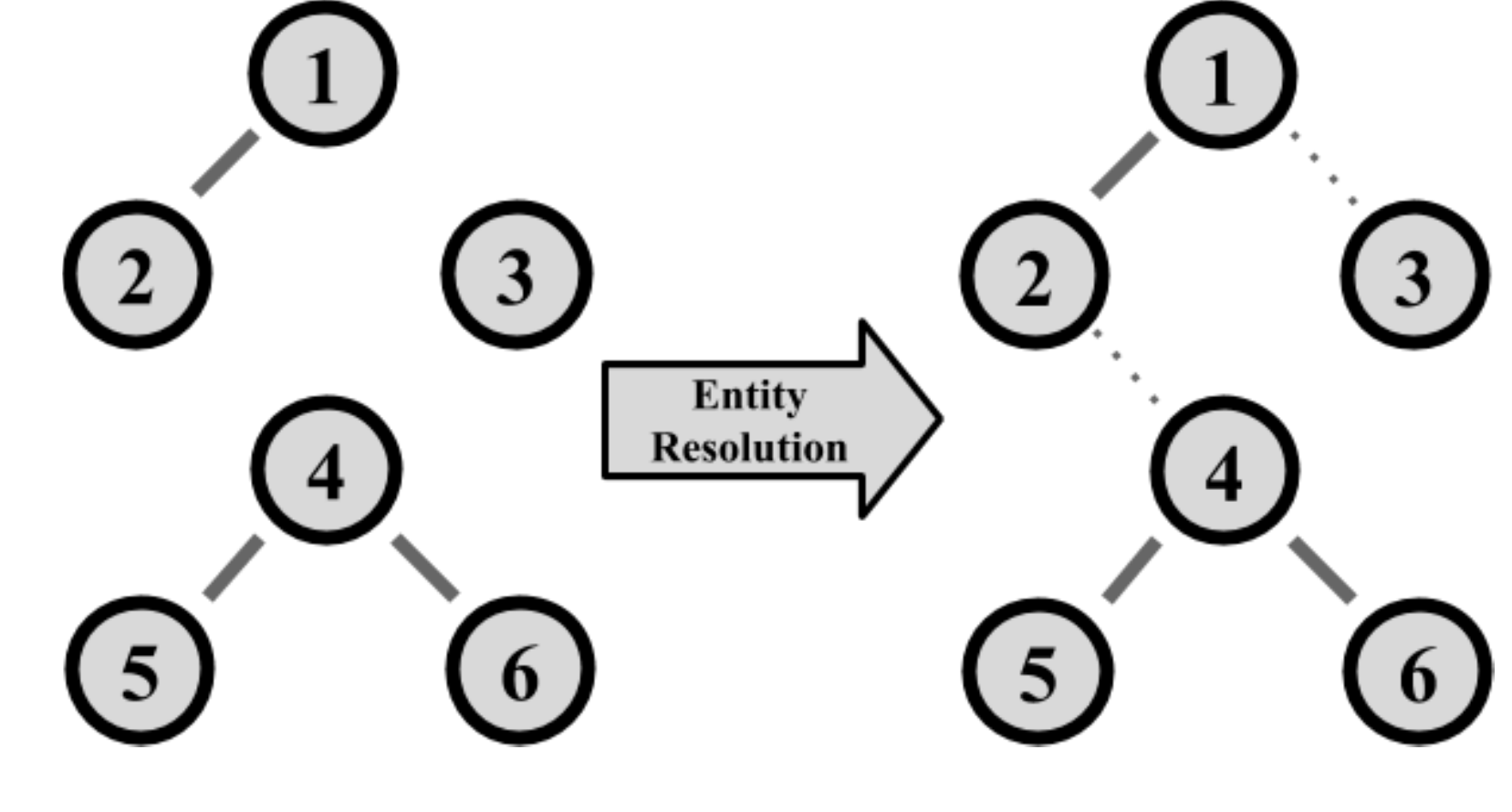}
 \caption{On applying our match function, weak links are generated for classifier scores above a certain match threshold. The strong links between nodes are represented by Solid Lines. Dashed lines represent the weak links generated by our classifier. }

\end{figure}

\subsection{Sampling Scheme}
For our classifier we need to generate a set of training examples `$\mathcal{T}$', and $\mathcal{T}_{pos}$ \& $\mathcal{T}_{neg}$ are the subsets of samples labeled positive and negative.\\
$\mathcal{T}_{pos} = \{ F_{v_{i},v_{j}} | v_{i} \in \mathcal{P}(p_{i})$, $v_{j}\in \mathcal{P}(p_{i}), \forall p_{i} \in S \}$ \\
$\mathcal{T}_{neg} = \{ F_{v_{i},v_{j}}| v_{i} \in \mathcal{P}(p_{i})$, $v_{j}\not\in \mathcal{P}(p_{i}), \forall p_{i} \in S \}$

\begin{figure}[!htbp]
\centering
\includegraphics[width=6cm]{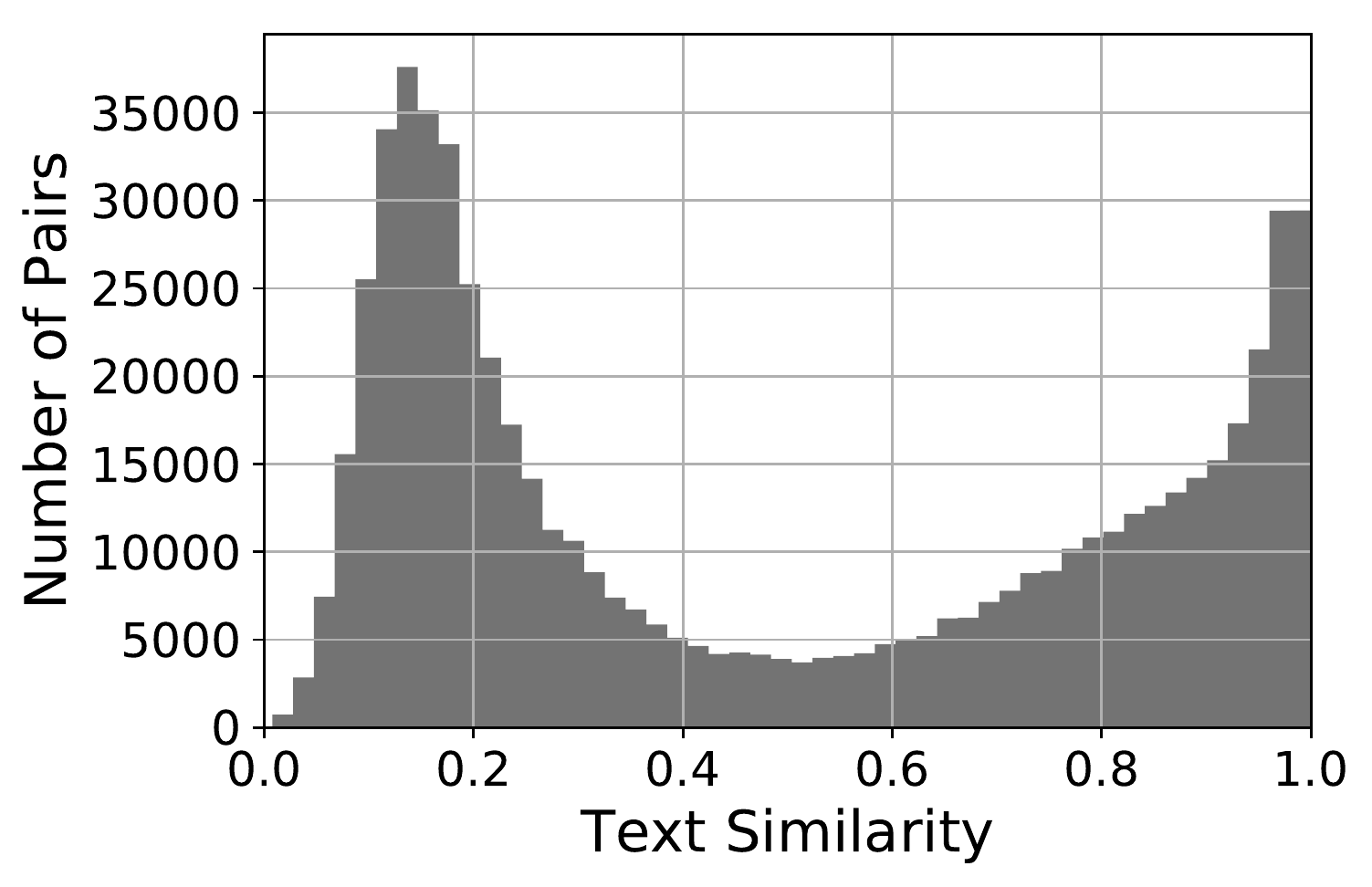}
\caption{Text Similarity for our Sampling Scheme. We use Jaccards Similarity between the ad unigrams as a measure of text similarity. The histogram shows that the sampling scheme results in both, a large number of near duplicates and non duplicates. Such a behavior is desired to ensure a robust match function.}
\label{fig:jaccards}
\end{figure}

In order to ensure that the sampling scheme does not end up sampling near duplicate pairs, we introduce a sampling bias such that for every feature vector $F_{v_{i},v_{j}} \in \mathcal{T}_{pos} $, $\mathcal{S}_{v_{i}} \cap \mathcal{S}_{v_{j}} = \phi$\\
This reduces the likelihood of sampling near-duplicates as evidenced in Figure \ref{fig:jaccards}, which is a histogram of the Jaccards Similarity  between the set of the unigrams of the text contained in the pair of ads. 

$sim(v_{i}, v_{j}) = \frac{|\mathtt{unigrams}(v_{i}) \cap \mathtt{unigrams}(v_{j})|}{|\mathtt{unigrams}(v_{i}) \cup \mathtt{unigrams}(v_{j})|}$\\
We observe that although we do still end with some near duplicates ($sim>0.9$), we have high number of non duplicates. ($0.1<sim<0.3$) which ensures robust training data for our classifier.

\begin{figure*}[!t]
\captionsetup[subfigure]{labelformat=empty}

\centering
    \begin{subfigure}{0.24\textwidth}
        \centering
        \includegraphics[width=4cm]{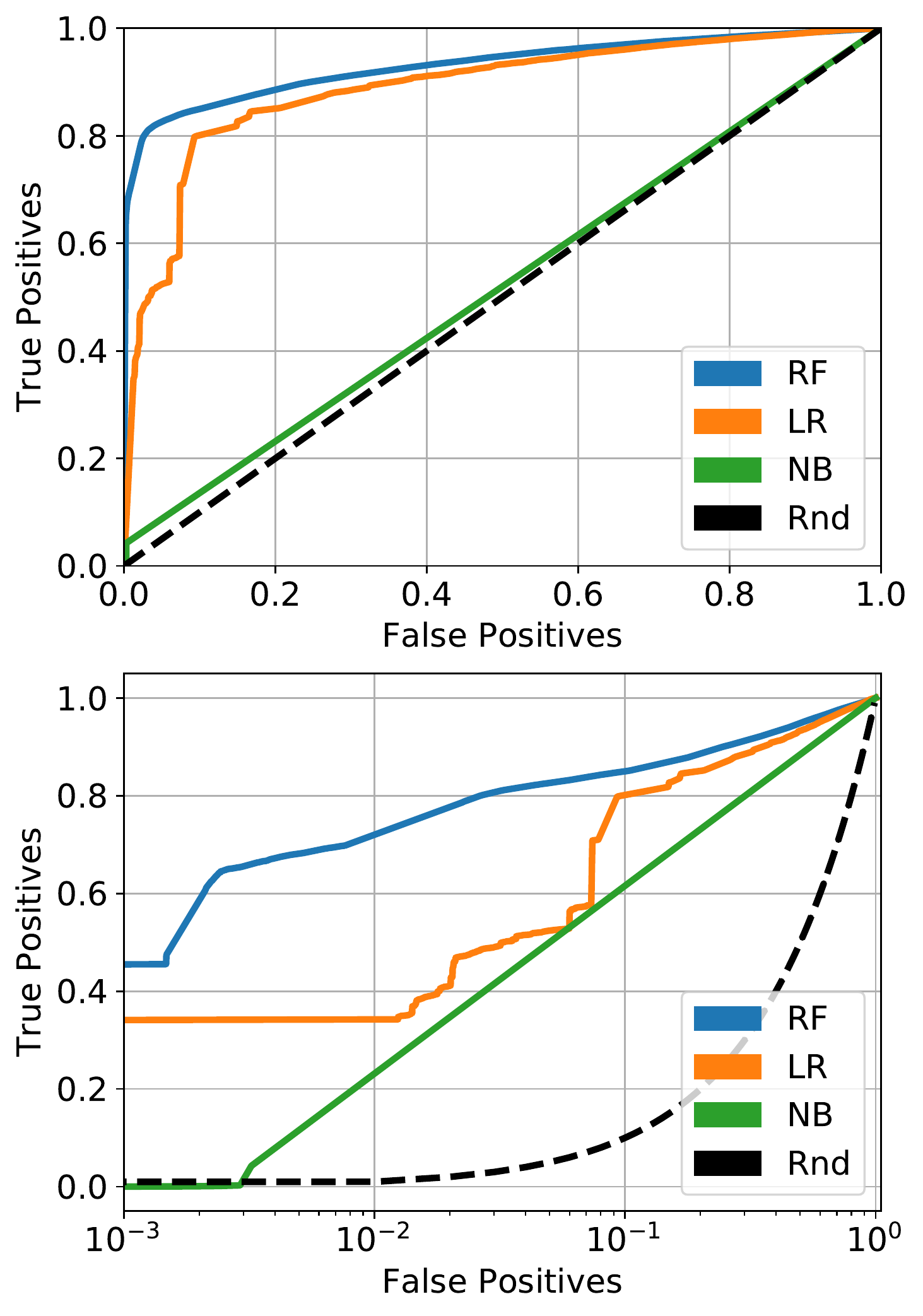}
        \caption{\small{\texttt{Regx}}}
    \end{subfigure}
    \begin{subfigure}{0.24\textwidth}
        \centering
        \includegraphics[width=4cm]{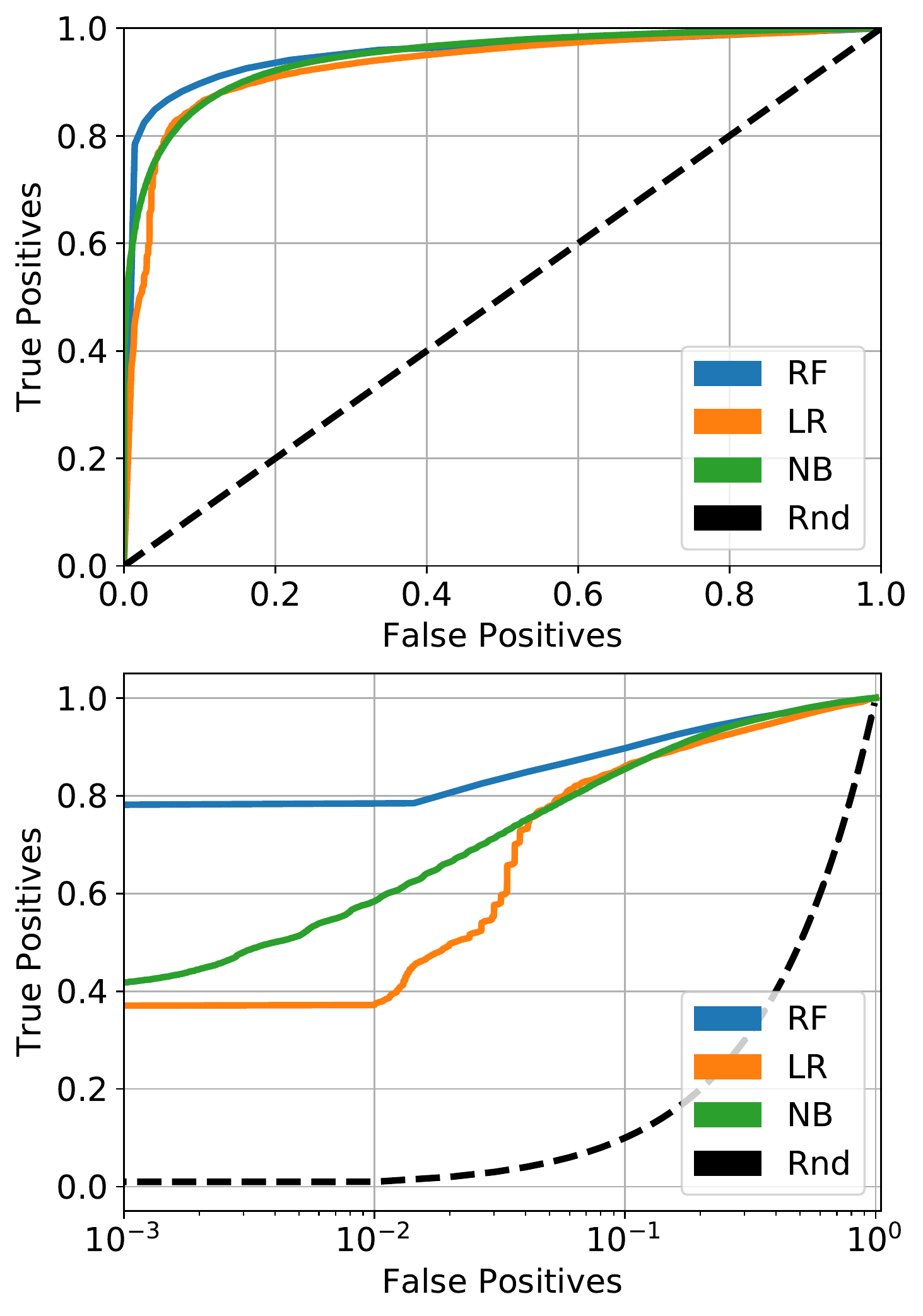}
        \caption{\small{\texttt{Regx+Temporal}}}
    \end{subfigure}
        \begin{subfigure}{0.24\textwidth}
        \centering
        \includegraphics[width=4cm]{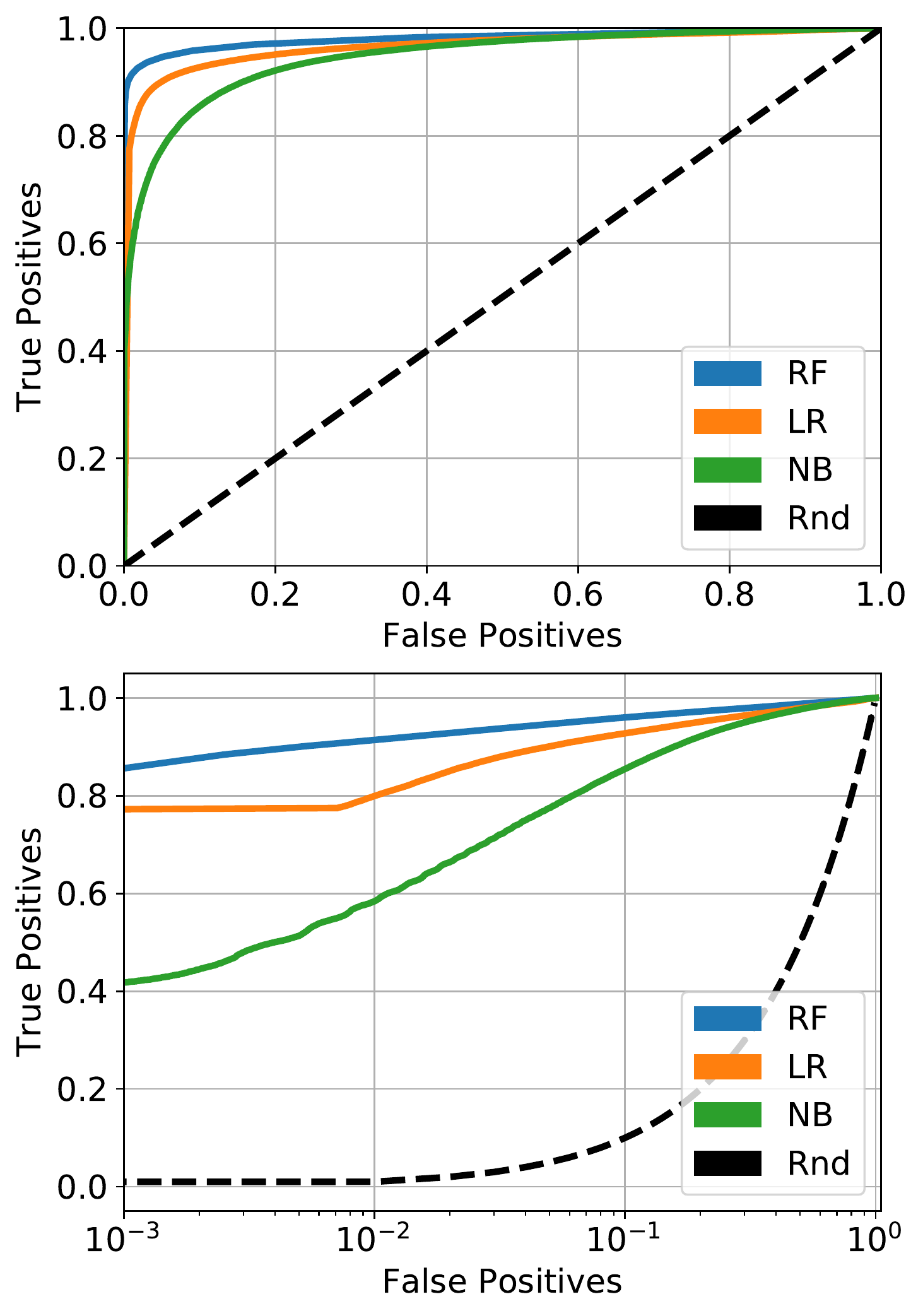}
        \caption{\small{\texttt{Regx+Temporal+NLP}}}
    \end{subfigure}
        \begin{subfigure}{0.24\textwidth}
        \centering
        \includegraphics[width=4cm]{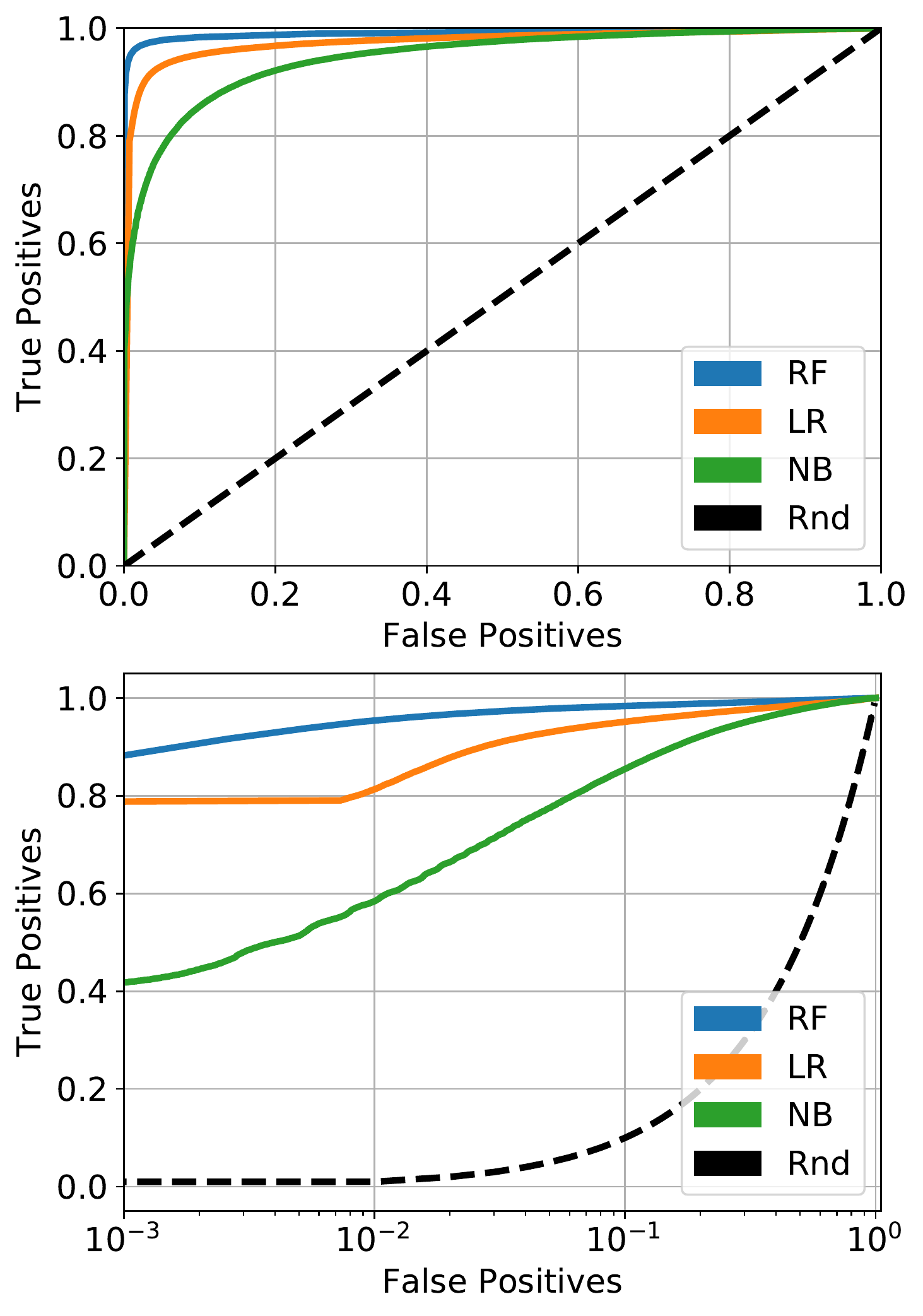}
        \caption{\tiny{\texttt{Regx+Temporal+NLP+Spatial}}}
    \end{subfigure}

\caption{ROC Curves for our Match Function trained on various feature sets. The ROC curve shows reasonably large True Positive rates for extremely low False Positive rates, which is a desirable behaviour of the match function.}
\label{plot:roc}
\end{figure*}

\subsection{Training}
To train our classifier we experiment with various classifiers like Logistic Regression, Naive Bayes and Random Forest using Scikit.~\cite{Pedregosa:2011:SML:1953048.2078195} Table \ref{tab:mosti1} shows the most informative features learnt by the Random Forest classifier. It is interesting to note that the most informative features include, the spatial (Location), Temporal (Time Difference, Posting Date) and also the Linguistic (Number of Special Characters, Longest Common Substring) features. We also find that the domain specific features, extracted using regexs, prove to be informative.

\begin{table}[!htbp]
\caption{Most Informative Features}
\label{tab:mosti1}
\centering
\begin{tabular}{|c|c|}
\hline
& \textbf{Top 10 Features} \\
\hline

\texttt{1} & \texttt{Location (State)} \\ \hline
\texttt{2} & \texttt{Number of Special Characters} \\ \hline
\texttt{3} & \texttt{Longest Common Substring} \\ \hline
\texttt{4} & \texttt{Number of Unique Tokens} \\ \hline
\texttt{5} & \texttt{Time Difference} \\ \hline
\texttt{6} & \texttt{If Posted on Same Day} \\ \hline
\texttt{7} & \texttt{Presence of Ethnicity} \\ \hline
\texttt{8} & \texttt{Presence of Rate} \\ \hline
\texttt{9} & \texttt{Presence of Restrictions} \\ \hline
\texttt{10} & \texttt{Presence of Names} \\ \hline

\end{tabular}
\end{table}

\begin{figure}[!htbp]
\captionsetup[subfigure]{labelformat=empty}
\centering
    \begin{subfigure}{0.245\textwidth}

        \centering
        \includegraphics[width=\linewidth]{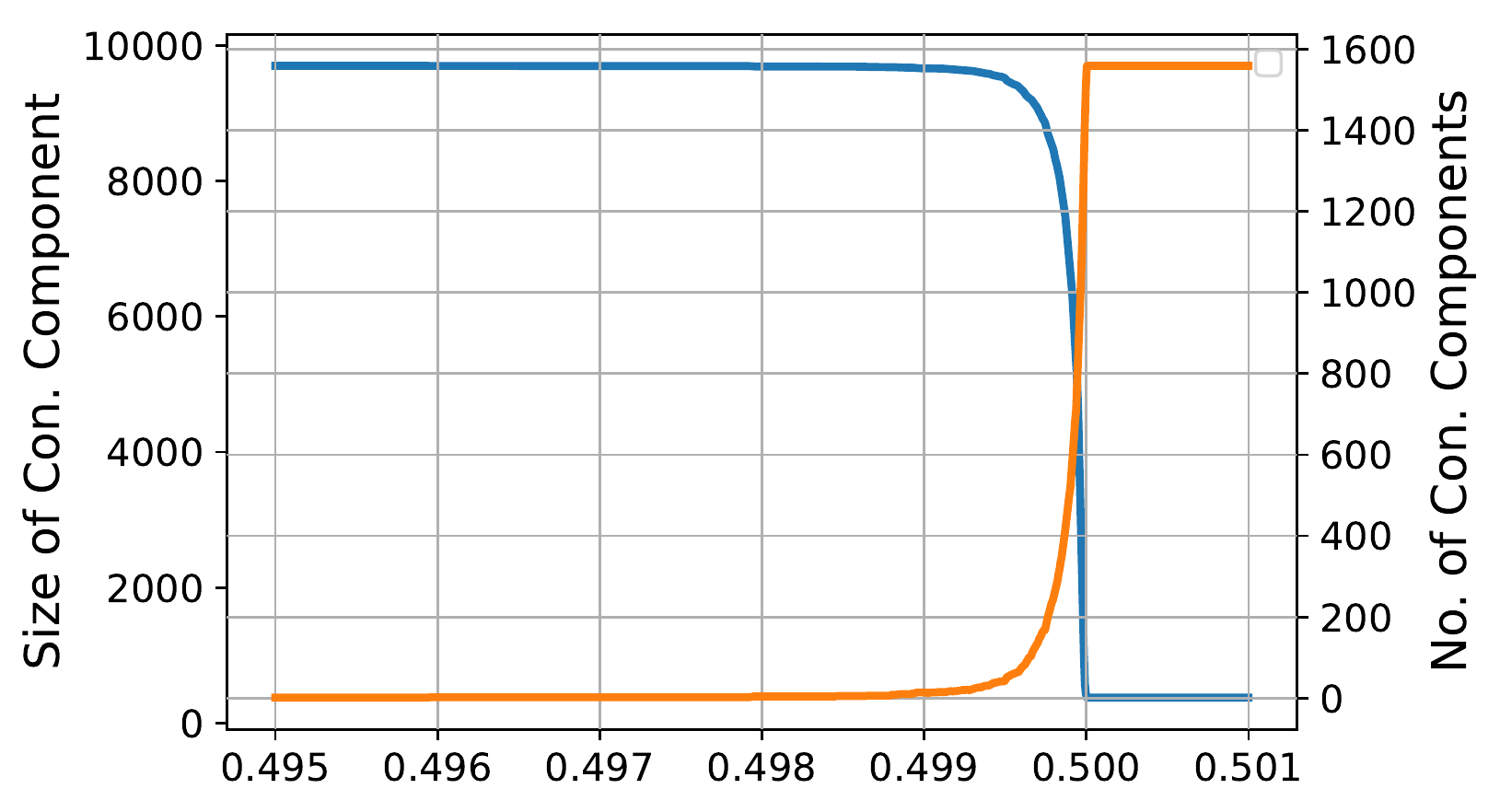}
        \caption{\texttt{Logistic Regression}}
    \end{subfigure} \hspace{-1.15em}
       \begin{subfigure}{0.245\textwidth}
        \centering
        \includegraphics[width=\linewidth]{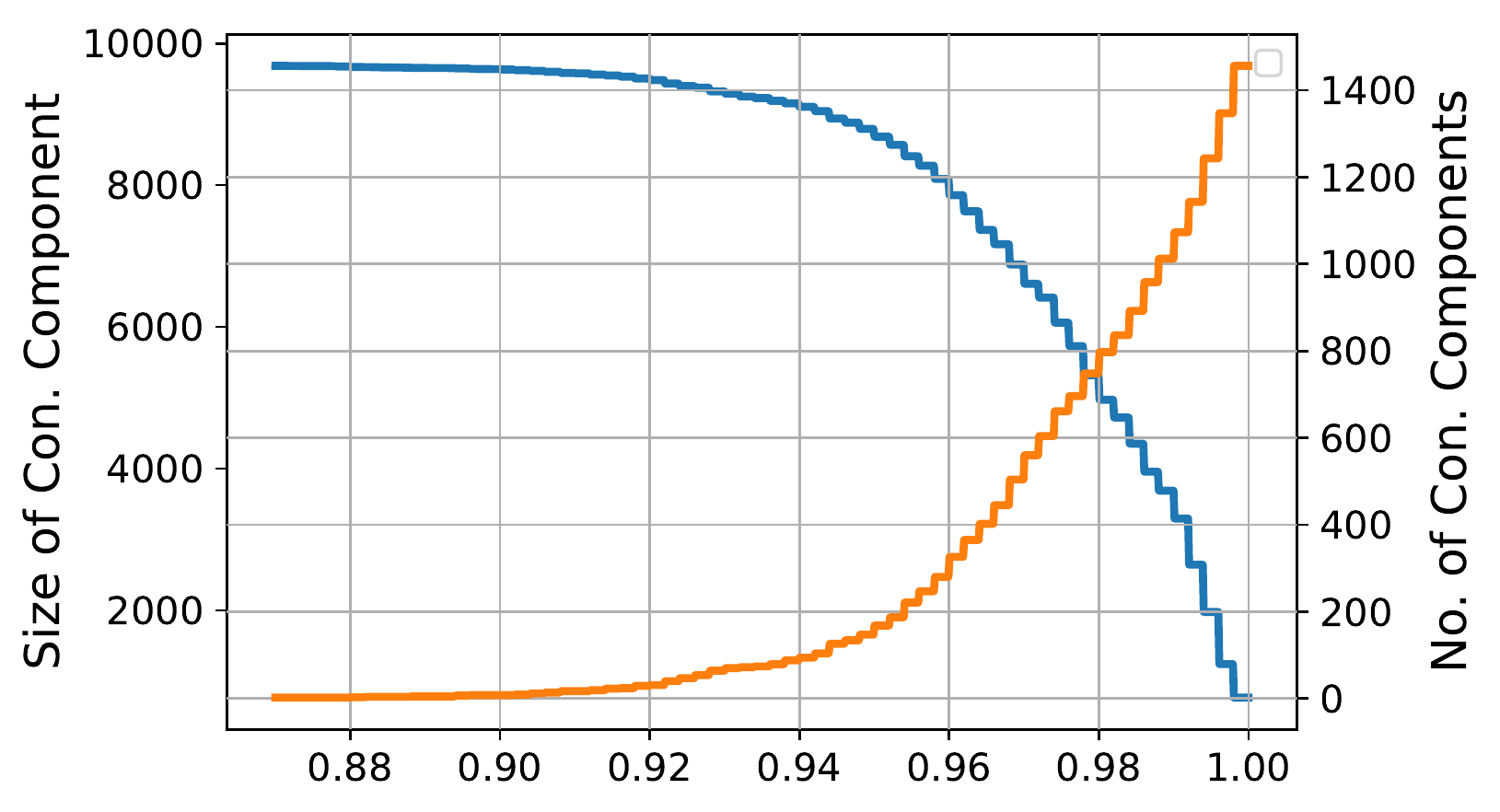}
        \caption{\texttt{Random Forest}}
    \end{subfigure}
\caption{The plots represents the number of connected components and the size of the largest component versus the match threshold.}
\label{plot:breakdown}
\end{figure}

The ROC curves for the classifiers we tested with different feature sets are presented in Figure \ref{plot:roc}. The classifiers performs well, with extremely low false positive rates. Such a behavior is desirable for the classifier to act as a match function, in order to generate sensible results for the downstream tasks. High False Positive rates, increase the number of links between our records, leading to a `snowball effect' which results in a break-down of the downstream Entity Resolution process as evidenced in Figure \ref{plot:breakdown}.

\begin{table*}[!htbp]
\small
\centering
\caption{Results Of Rule Learning}
\label{tab:rules}
\centerline{\begin{tabular}{|c|c|c|c|}
\hline
\textbf{Rule} & \textbf{Support} & \textbf{Ratio} & \textbf{Lift}\\
\hline
\texttt{Xminchars<=250, 120000<Xmaximgfrq, 3<Xmnweeks<=3.4, 4<Xmnmonths<=6.5} & 11 & 90.9\% & 2.67 \\ \hline
\texttt{  Xminchars<=250, 120000<Xmaximgfrq 4<Xmnmonths<=6.5, 
} & 16 & 81.25\% &2.4\\ \hline
\texttt{    Xstatesnorm<=0.03, 3.6<Xuniqimgsnorm<=5.2, 3.2<Xstdmonths} & 17 & 100.0\% &2.5\\ \hline
\texttt{Xstatesnorm<=0.03, 1.95<Xstdweeks<=2.2, 3.2<Xstdmonths} & 19 &94.74\% &2.37\\ \hline
\end{tabular}}
\end{table*}

In order to minimize this breakdown, we need to heuristically learn an appropriate confidence value for our classifier. This is done by carrying out the ER process on 10,000 randomly selected records from our dataset. The value of size of the largest extracted connected component and the number of such connected components isolated is calculated for different confidence values of our classifier. This allows us to come up with a sensible heuristic for the confidence value.

\begin{figure}[!htbp]

\captionsetup[subfigure]{labelformat=empty}

\centering
    \hspace{-2.2em}
    \begin{subfigure}{0.16\textwidth}
        \centering
        \includegraphics[width=1.2\linewidth]{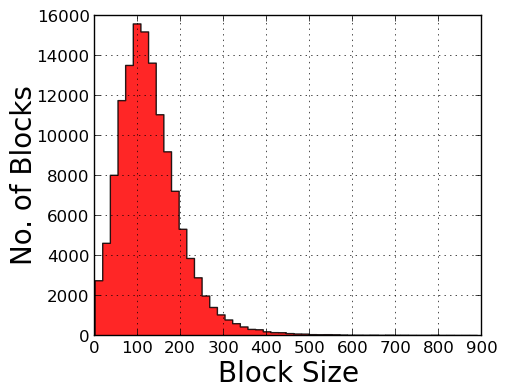}
        \caption{\hspace{3em}\texttt{Bigrams}}
    \end{subfigure}
    \hfill
    \begin{subfigure}{0.16\textwidth}
        \centering
        \includegraphics[width=1.2\linewidth]{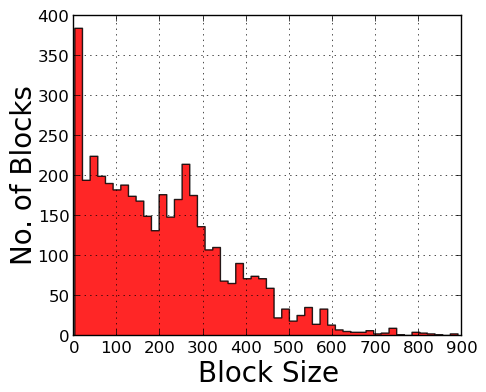}
        \caption{\hspace{3em}\texttt{Unigrams}}
    \end{subfigure}
    \hfill
    \begin{subfigure}{0.16\textwidth}
        \centering
        \includegraphics[width=1.2\linewidth]{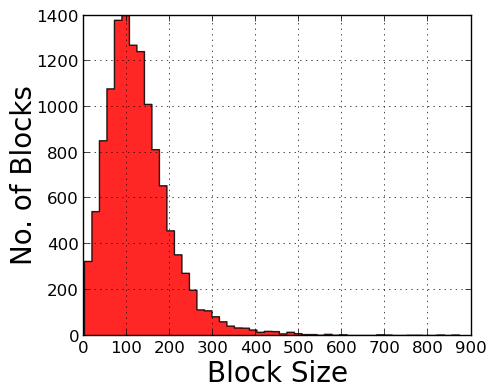}
        \caption{\hspace{3em}\texttt{Images}}
    \end{subfigure}

\caption{Blocking Scheme}
\end{figure}

\subsection{Blocking Scheme}
Our dataset consists of over 5 million records. Naive pairwise comparisons across the dataset, makes this problem computationally intractable. In order to reduce the number of comparisons, we introduce a blocking scheme and performa exhaustive pairwise comparisons only within each block before resolving the dataset across blocks. We block the dataset on features like Rare Unigrams, Rare Bigrams and Rare Images.

\begin{figure}[!htbp]
\centering
    \begin{subfigure}[b]{0.5\textwidth}
        \centering
        \shadowbox{\includegraphics[width=0.8\textwidth]{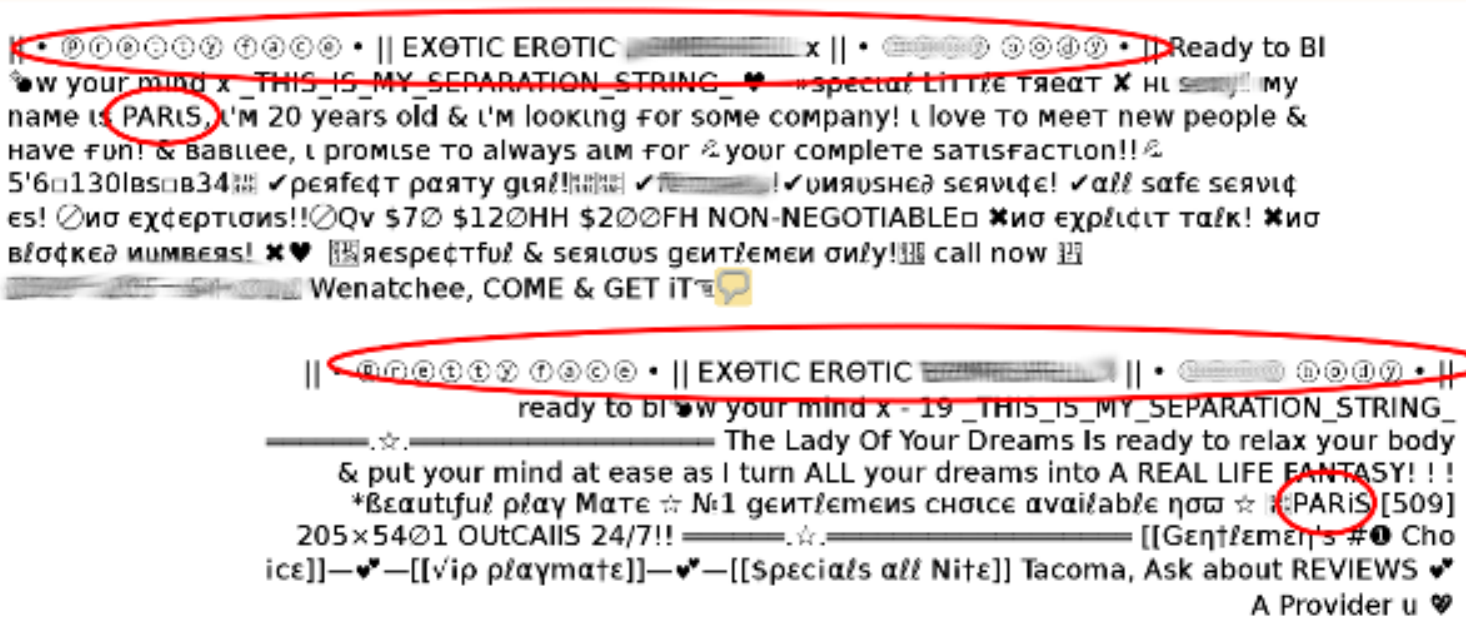}}
        \caption{This pair of ads have extremely similar textual content including use of non-latin and special characters. The ad also advertises the same individual, as strongly evidenced by the common alias, `Paris'.}
        \vspace{1.5em}
    \end{subfigure}
      \begin{subfigure}[b]{0.5\textwidth}
        \centering
        \shadowbox{\includegraphics[width=0.8\textwidth]{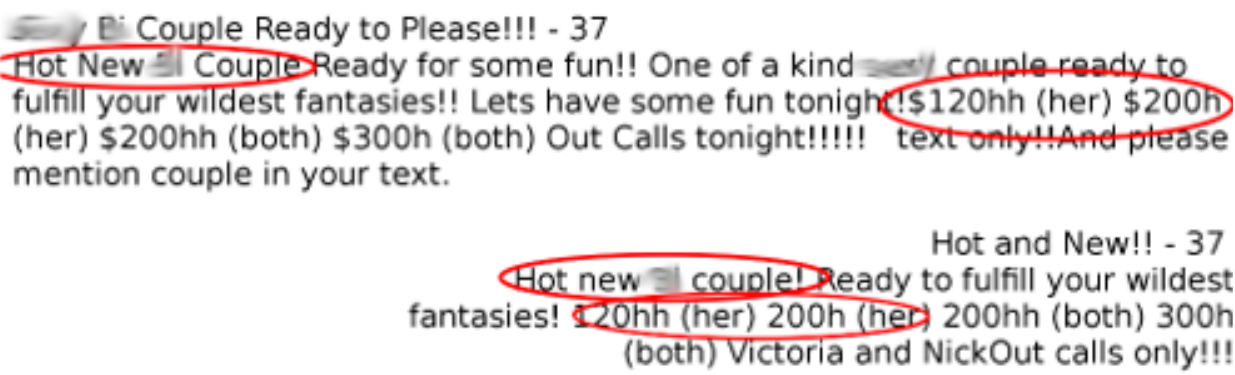}}
        \caption{The first ad here does not include any specific names of individuals. However, The strong textual similarity with the second ad and the same advertised cost, helps to match them and discover the individuals being advertised as `Nick' and `Victoria'.}
        \vspace{1.5em}

    \end{subfigure}
    \begin{subfigure}[b]{0.5\textwidth}
        \centering
        \shadowbox{\includegraphics[width=0.8\textwidth]{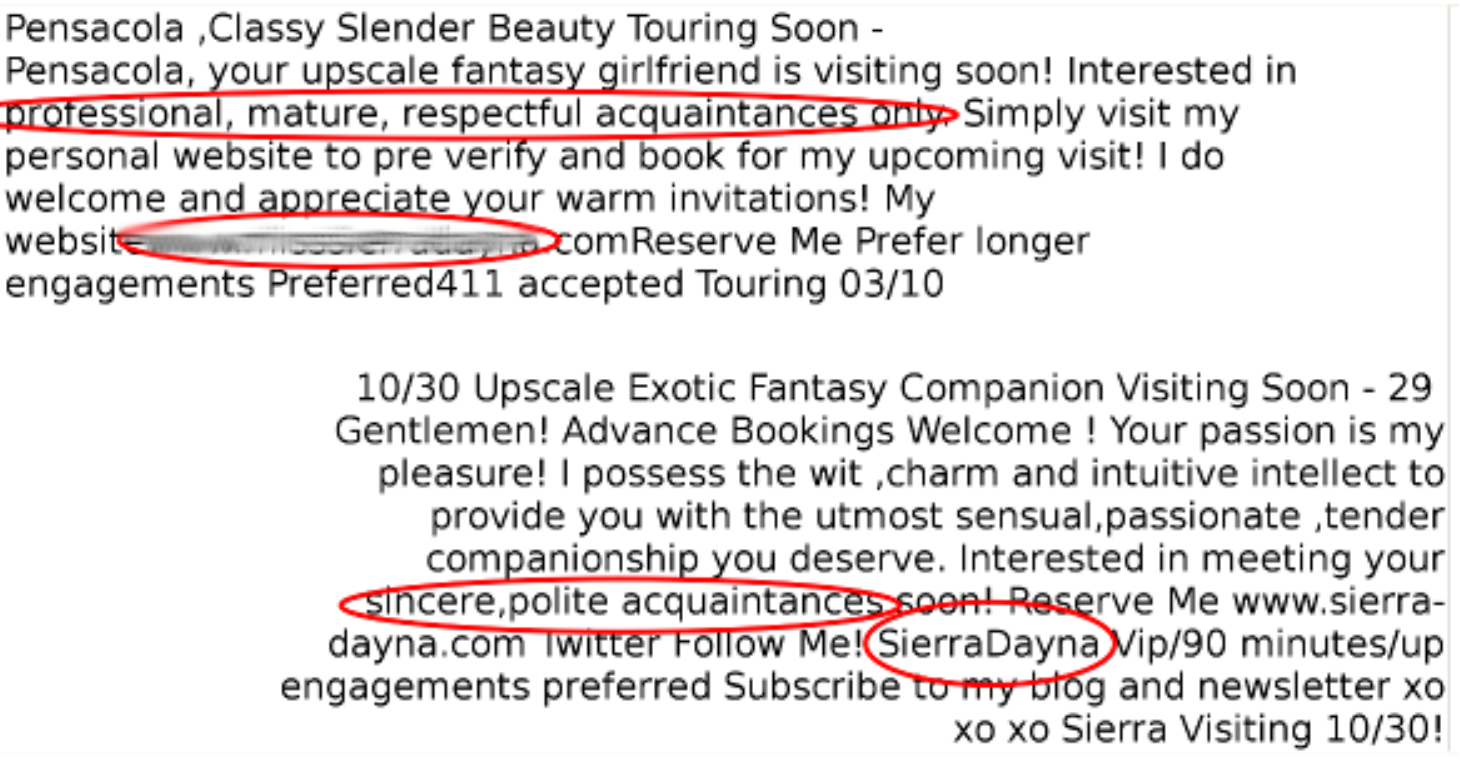}}
        \caption{While this pair is not extremely similar in terms of language, however the existence of the rare alias `SierraDayna' in both advertisemets helps the classifier in matching them. This match can also easily be verified by the similar language structure of the pair.}
        \vspace{1.5em}
    \end{subfigure}
    \begin{subfigure}[b]{0.5\textwidth}
        \centering
        \shadowbox{\includegraphics[width=0.8\textwidth]{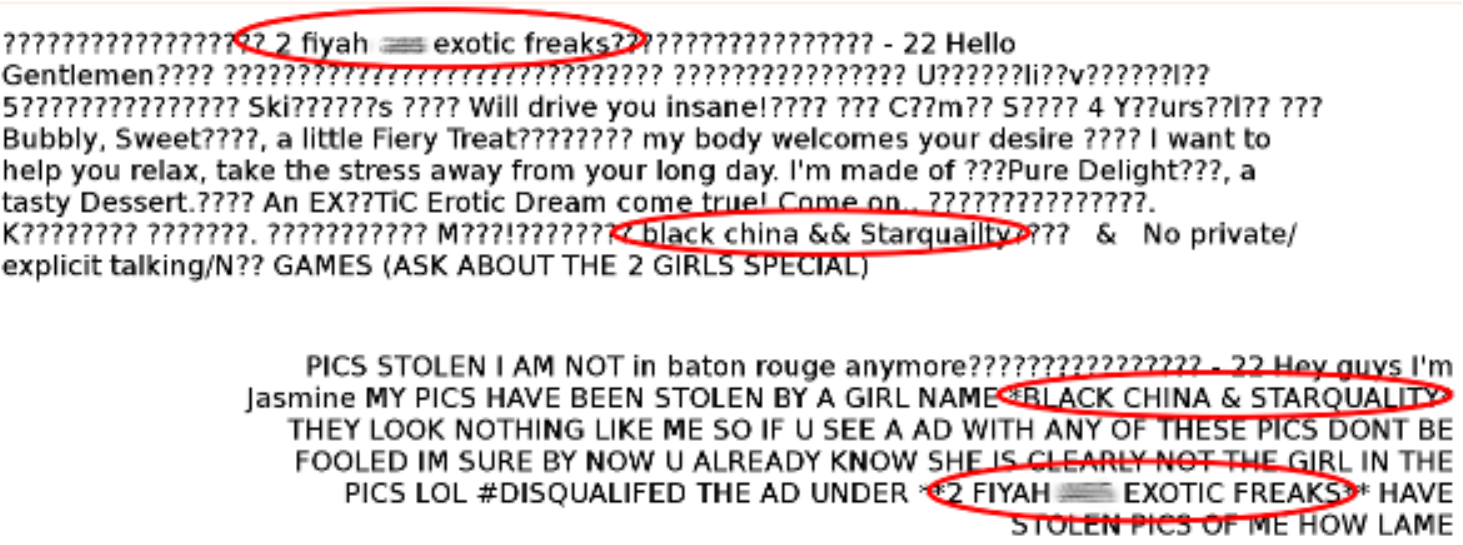}}
        \caption{The first advertisement represents entities `Black China' and `Star Quality', while the second advertisement, reveals that the pictures used in the first advertisement are not original and belong to the author of the second ad. This example pair shows the robustness of our match function. It also reveals how complicated relationships between various ads can be.}
        \vspace{1em}

    \end{subfigure}
    
\caption{Representative results of advertisement pairs matched by our classifier. In all the four cases the advertisement pairs had no phone number information (strong feature) in order to detect connections. Note that sensitive elements have been intentionally obfuscated.}
\end{figure}

\begin{figure}[H]
\centering
\includegraphics[width=6cm]{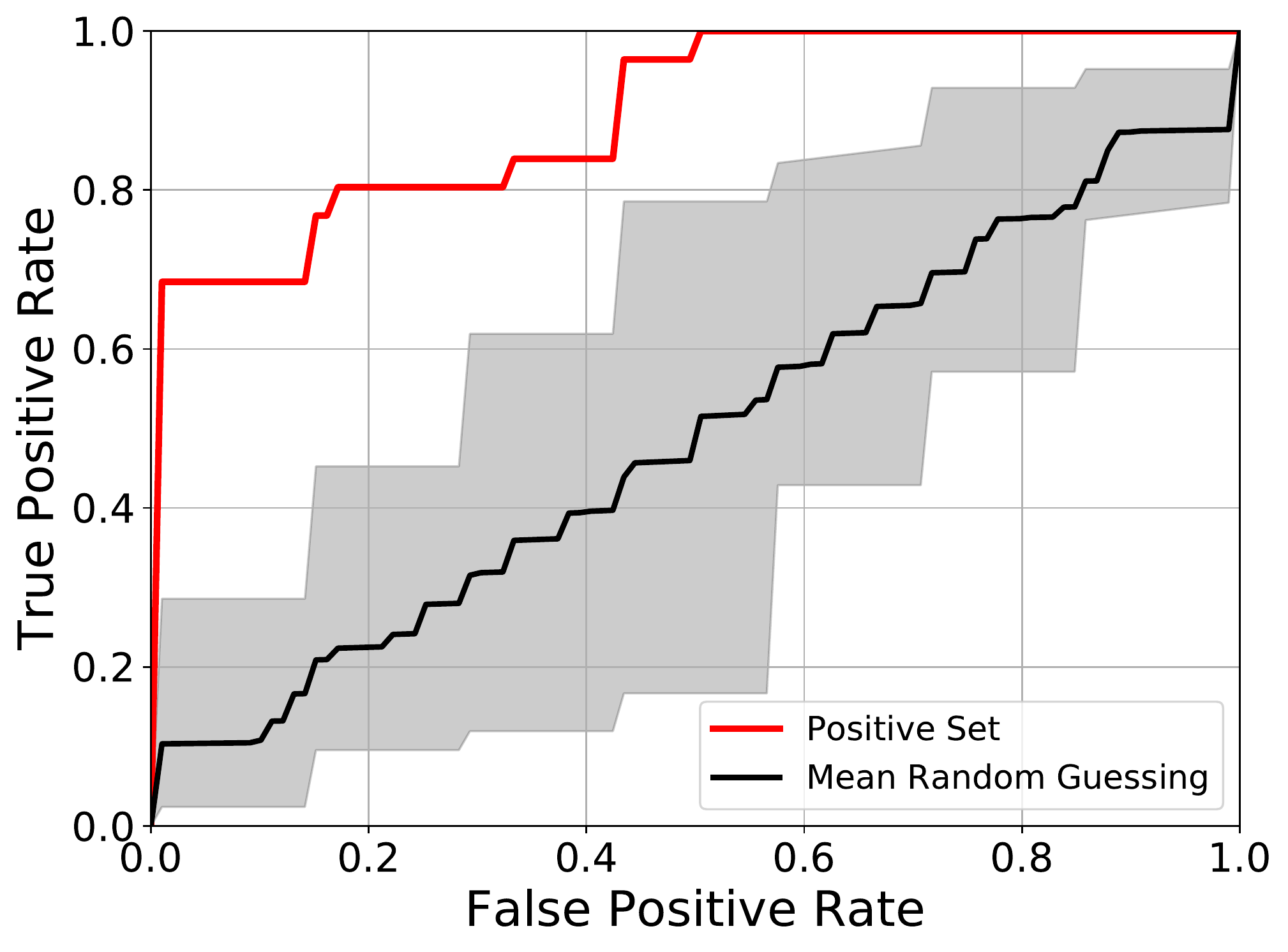}
\caption{ROC for the Connected Component classifier. The Black line is the positive set, while the Red line is the average ROC for 100 randomly guessed predictors.}
\label{plot:cc}
\end{figure}

\begin{figure}[H]
\centering
\includegraphics[width=6cm]{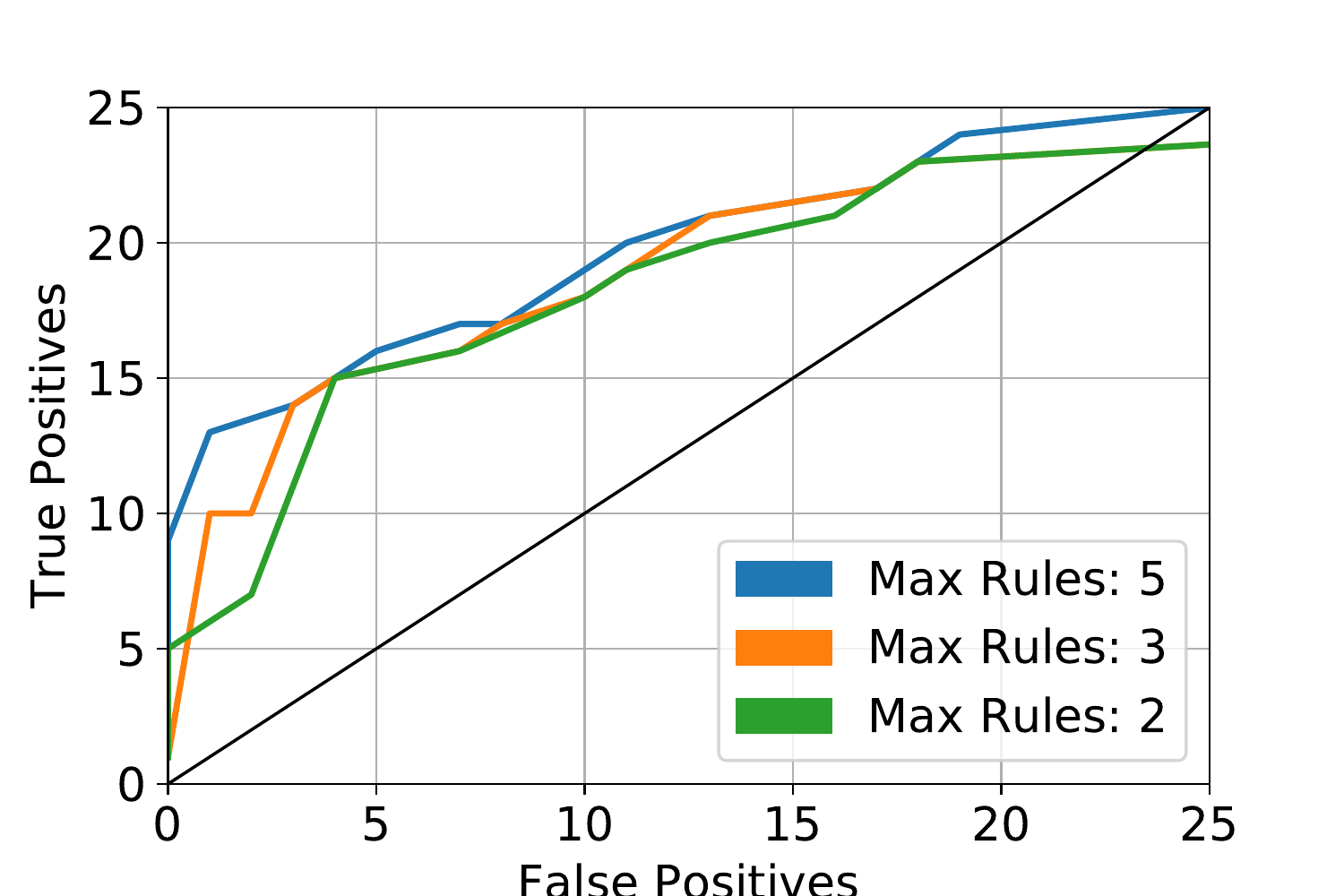}
\caption{PN Curve for rule learning. The figure presents PN curves for various values of the Maximum Rules learnt for the classification.}
\label{plot:pn}
\end{figure}

\section{Rule Learning}

We extract clusters and identify records that are associated with human trafficking using domain knowledge from experts. We featurize the extracted components, using features like size of the cluster, the spatio-temporal characteristics, and the connectivity of the clusters. For our analysis, we consider only components with more than 300 advertisements. we then train a random forest to predict if the clusters is linked to human trafficking. In order to establish statistical significance, we compare the ROC results of our classifier in 4 cross validation for 100 random connected components versus the positive set. 
\mbox{Figure \ref{plot:cc}} \& \mbox{Table \ref{tab:fprtpr}} lists the performance of the classifier in terms of False Positive and True Positive Rate while \mbox{Table \ref{tab:conncompfeats}} lists the most informative features for this classifier.

We then proceed to learn rules from our featureset. Some of the rules with corresponding Ratios and Lift are given in \mbox{Table \ref{tab:rules}}. PN curves corresponding to various rules learnt are presented in the Figure \ref{plot:pn} It can be observed that the features used by the rule learning to learn rules with maximum support and ratios, correspond to the ones labeled by the random forest as informative. This also serves as validation for the use of rule learning.

\begin{figure}[!t]
\centering
\centerline{\includegraphics[width=1.2\linewidth]{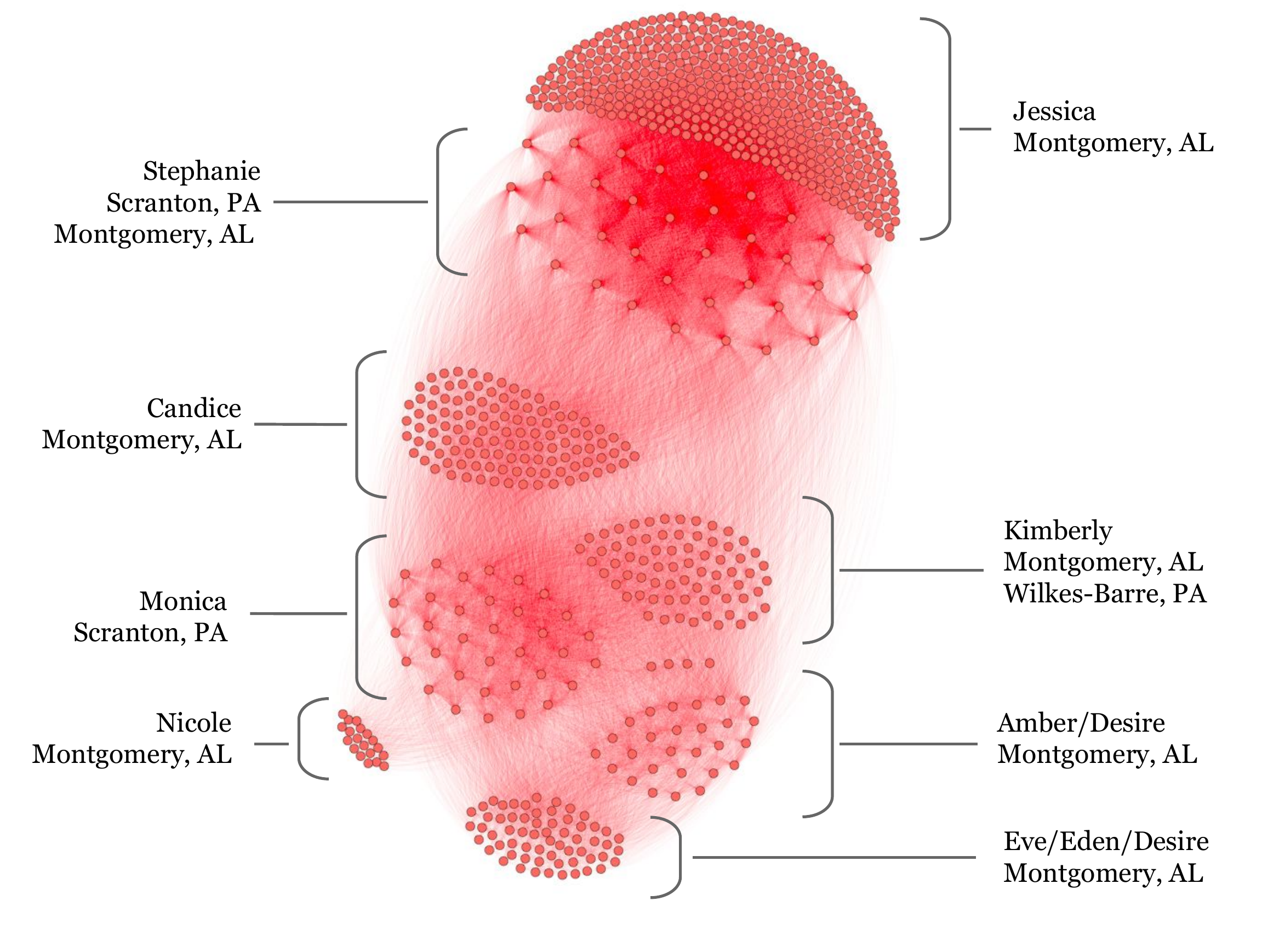}}
\caption{Representative Entity isolated by our pipeline, believed to be involved in human trafficking. The nodes represent advertisements, while the edges represent links between advertisements. This entity has 802 nodes and 39,383 edges. This visualization is generated using Gephi. \cite{ICWSM09154}. This entity operated in cities, across states and advertised multiple different individuals along with multiple phone numbers. This suggests a more complicated and organised activity and serves as an example of how complicated certain entities can be in this trade.}. 
\end{figure}

\begin{table}[!htbp]
\caption{Metrics for the Connected Component classifier}
\label{tab:fprtpr}
\centering
\begin{tabular}{|c|c|c|}
\hline
\textbf{AUC}& \textbf{TPR@FPR=1\%} & \textbf{FPR@TPR=50\%} \\
\hline

90.38\% & 66.6\% & 0.6\% \\ \hline
\end{tabular}
\end{table}

\begin{table}[!htbp]
\caption{Most Informative Features}
\label{tab:conncompfeats}
\centering
\begin{tabular}{|c|c|}
\hline
& \textbf{Top 5 Features} \\
\hline

\texttt{1} & \texttt{Posting Months} \\ \hline
\texttt{2} & \texttt{Posting Weeks} \\ \hline
\texttt{3} & \texttt{Std-Dev. of Image Frequency} \\ \hline
\texttt{4} & \texttt{Norm. No. of Names} \\ \hline
\texttt{5} & \texttt{Norm. No. of Unique Images} \\ \hline

\end{tabular}
\end{table}

\section{Conclusion}
In this paper we approached the problem of isolating sources of human trafficking from online escort advertisements with a pairwise Entity Resoltuion approach. We trained a classifier able to predict if two advertisements are from the same source using phone numbers as a strong feature and exploit it as proxy ground truth to generate training data for our classifier. The resultant classifier, proved to be robust, as evidenced from extremely low false positive rates. Other approraches \cite{szekely15-iswc} aims to build similar knowledge graphs using similarity score between each feature. This has some limitations. Firstly, we need labelled training data inorder to train match functions to detect ontological relations. The challenge is aggravated since this approach considers each feature independently making  generation of enough labelled training data for training multiple match functions an extremely complicated task.

Since we utilise existing features as proxy evidence, our approach can generate a large amount of training data without the need of any human annotation. Our approach requires just learning a single function over the entire featureset, hence our classifier can learn multiple complicated relations between features to predict a match, instead of the naive feature independence assumption. 

We then proceeded to use this classifier in order to perform entity resolution using a heurestically learned value for the score of classifier, as the match threshold.
The resultant connected components were again featurised, and a classifier model was fit before subjecting to rule learning. On comparison with \cite{dubrawski2015leveraging}, the connected component classifier performs a little better with higher values of the area under the ROC curve and the TPR@FPR=1\% indicating a steeper, ROC curve. We hypothesize that  due to the entity resolution process, we are able to generate larger, more robust amount of training data which is immune to the noise in labelling and results in a stronger classifier. The learnt rules show high ratios and lift for reasonably high supports as evidenced from Table \ref{tab:rules}. Rule learning also adds an element of interpretability to the models we built, and as compared to more complex ensemble methods like Random Forests, having hard rules as classification models are preferred by Domain Experts to build evidence for incrimination.

\section{Future Work}
While our blocking scheme performs well to reduce the number of comparisons, however since our approach involves naive pairwise comparisons, scalability is a significant challenge. One approach could be to design such a pipeline in a distributed environment. Another approach could be to use a computationally inexpensive technique to de-duplicate the dataset of the near duplicate ads, which would greatly help with regard to scalability.

In our approach, the ER process depends upon the heuristically learnt match threshold. Lower threshold values can significantly degrade the performance, with extremely large connected components. The possibility of treating this attribute as a learning task, would help making this approach more generic, and non domain specific.

Hashcodes of the images associated with the ads were also utilized as a feature for the match function. However, simple features like number of unique and common images etc., did not prove to be very informative. Further research is required in order to make better use of such visual data.


\section*{Acknowledgments}

The authors would like to thank all staff, faculty and students who made the Robotics Institute Summer Scholars program 2015 at Carnegie Mellon University possible.

\bibliography{emnlp2017}
\bibliographystyle{emnlp_natbib}

\end{document}